\documentclass[11pt,usenatbib]{mnras}

\usepackage{graphicx}
\usepackage{caption}
\usepackage{epstopdf}
\usepackage{amsmath}
\usepackage{wasysym}
\usepackage{lmodern}
\usepackage{slantsc}

\def\gs{\mathrel{\raise0.35ex\hbox{$\scriptstyle >$}\kern-0.6em
\lower0.40ex\hbox{{$\scriptstyle \sim$}}}}
\def\ls{\mathrel{\raise0.35ex\hbox{$\scriptstyle <$}\kern-0.6em
\lower0.40ex\hbox{{$\scriptstyle \sim$}}}}
\def\ls{\mathrel{\hbox{\rlap{\hbox{\lower4pt\hbox{$\sim$}}}\hbox{$<$}}}}
\def\gs{\mathrel{\hbox{\rlap{\hbox{\lower4pt\hbox{$\sim$}}}\hbox{$>$}}}}
																																	
\def\mnras {{\sc MNRAS}}
\def\apj {ApJ}
\def\apjs {ApJS}
\def\apjl {ApJL}
\def\aj {AJ}

\def\pasp {PASP}

\newcommand{\ur}{\mbox{$(u-r)$}}
\newcommand{\nuvr}{\mbox{$({\rm NUV} - r)$}}
\newcommand{\nuvu}{\mbox{$({\rm NUV} - u)$}}

\newcommand\gtsim{\mathrel{\lower0.6ex\hbox{$\buildrel {\textstyle >}
  \over {\scriptstyle \sim}$}}}
\newcommand\ltsim{\mathrel{\lower0.6ex\hbox{$\buildrel {\textstyle <}
  \over {\scriptstyle \sim}$}}}

\newcommand{\degree}{$^{\circ}$ }

\title[]
      {NUV signatures of environment driven galaxy quenching in SDSS groups}
\author[Crossett, Pimbblet, Jones, Brown, \& Stott]
       {Jacob P.\ Crossett$^{1,2,3}$\thanks{email: jake.crossett@monash.edu},
        Kevin A.\ Pimbblet$^{1,3}$,
        D. Heath Jones$^{1,4,5}$,
       \and Michael J.\ I.\ Brown$^{1,2}$,
        John P.\ Stott$^{6}$
        \vspace*{1mm}\\
        $^{1}$School of Physics and Astronomy, Monash University, Clayton, Victoria 3800, Australia\\
        $^{2}$Monash Centre for Astrophysics (MoCA), Monash University, Clayton, Victoria 3800, Australia\\
        $^{3}$E. A. Milne Centre for Astrophysics, University of Hull, Cottingham Road, Kingston-upon-Hull, HU6 7RX, U.K.\\
        $^{4}$English Language and Foundation Studies Centre, University of Newcastle, Callaghan NSW 2308, Australia\\
        $^{5}$Department of Physics and Astronomy, Macquarie University, Sydney NSW 2109, Australia\\
        $^{6}$Department of Physics, University of Oxford, Keble Road, Oxford, OX1 3RH, U.K.\\ }

\pagerange{000--000}

\begin{document}

\maketitle
\begin{abstract}
We have investigated the effect of group environment on residual star formation in galaxies, using \textsl{\textsc{Galex}} NUV galaxy photometry with the SDSS group catalogue of \citet[][]{Yang:2007aa}. We compared the \nuvr\ colours of grouped and non-grouped galaxies, and find a significant increase in the fraction of red sequence galaxies with blue \nuvr\ colours outside of groups. When comparing galaxies in mass matched samples of satellite (non-central), and non-grouped galaxies, we found a $>4\sigma$ difference in the distribution of \nuvr\ colours, and an \nuvr\ blue fraction $>3\sigma$ higher outside groups. A comparison of satellite and non-grouped samples has found the NUV fraction is a factor of $\sim2$ lower for satellite galaxies between $10^{10.5}M_{\astrosun}$ and $10^{10.7}M_{\astrosun}$, showing that higher mass galaxies are more able to form stars when not influenced by a group potential. There was a higher \nuvr\ blue fraction of galaxies with lower S\'ersic indices ($n < 3$) outside of groups, not seen in the satellite sample. We have used stellar population models of \citet[][]{Bruzual:2003aa} with multiple burst, or exponentially declining star formation histories to find that many of the \nuvr\ blue non-grouped galaxies can be explained by a slow ($\sim2$ Gyr) decay of star formation, compared to the  satellite galaxies. We suggest that taken together, the difference in \nuvr\ colours between samples can be explained by a population of secularly evolving, non-grouped galaxies, where star formation declines slowly. This slow channel is less prevalent in group environments where more rapid quenching can occur. 
\end{abstract}

\begin{keywords}
galaxies: evolution - galaxies: stellar content - galaxies: groups: general - galaxies: star formation - ultraviolet: galaxies - galaxies: photometry
\end{keywords}
\section{Introduction}

Large galaxy surveys have revealed the colour bimodality of galaxy populations, with the evolution of stellar masses in each population suggesting that these galaxies transition between a blue and red population. Colour bimodality is a generalisation of the result of \citet[][]{Visvanathan:1977aa} who found elliptical galaxies in a tightly constrained colour--magnitude red sequence, with only small metallicity and age differences \citep[][]{Kodama:1998aa}. While initially found in clusters \citep[e.g.][and references therein]{Bower:1992aa}, this red sequence has been seen in populations of galaxies in large surveys \citep[e.g.][]{Blanton:2003aa, Baldry:2004aa}. Conversely, a star forming population is seen in the so called `blue cloud' -- a lower mass, broadly blue area of the colour--magnitude diagram, where young stars dominate the flux of a galaxy. It is common to use the optical colour of a galaxy to discriminate between galaxies with different mean stellar ages, and likely different star formation histories \citep[e.g.][]{Strateva:2001aa, Hogg:2003aa, Baldry:2004aa, Balogh:2004aa, Stott:2007aa, van-den-Bosch:2008aa}.

While the optical bands give an indication of galaxy mean stellar age, alternative wavelengths can improve measurements of the current and recent star formation of a galaxy. One of the most direct measures of star formation comes from using ultraviolet light (UV), which is known to directly trace even low levels of star formation \citep[$<1M_{\astrosun}/\rm{year}$,][]{Salim:2007aa} using stars of a lifetime of $10^{8.5}$ years  \citep[][]{Martin:2005aa}. Even in the absence of evidence of star formation in optical photometry, the near UV (NUV) band can detect young stellar populations that have formed  within $\sim1$ Gyr \citep[e.g.][]{Ferreras:2000aa, Yi:2005aa}. This sensitivity to low levels of star formation causes the NUV-optical colour--magnitude diagram to lose the clear bimodal structure seen in optical colour--magnitude diagrams \citep[][]{Wyder:2007aa}. Instead a third population, the green valley, resides between the larger red sequence and blue cloud, suggesting differences in star formation levels unable to be distinguished in optical photometry \citep[][]{Wyder:2007aa, Salim:2014aa}. Galaxies in the green valley have shown intermediate specific star formation rates (log(sSFR)$\sim-11$), red optical colours, and intermediate masses ($M_{\astrosun} \sim 10^{10.8}$), indicating that they are in the process of moving between the blue cloud and the red sequence \citep[][]{Salim:2007aa, Schiminovich:2007aa, Salim:2014aa}. It is this breaking of the optical colour bimodality that shows several different stages of evolution may be present within a single broad colour population. 

One example of the several stages of evolution within a single broad population comes from \citet[][]{Schawinski:2014aa}, who found that splitting the green valley by morphological type will separate different transition paths. The green valley can be separated into two evolutionary paths, with elliptical galaxies predominantly found at the low mass end, and spirals at the high mass end. This indicates that the green valley contains two populations: one dominated by ellipticals, having ceased star formation quickly (a rapid quench, with a characteristic time of $<250$ Myr) and joining the red sequence at low mass; the other containing slowly evolving discs, which are undergoing a secular quenching of star formation (i.e. with times much greater than a dynamical timescale, which we define as $>1$Gyr), and joining the red sequence at higher masses \citep[][]{Kormendy:2004aa}.

The rapid quenching may have a similar evolution to early-type galaxies with signs of recent star formation \citep[][]{Yi:2005aa}. Such galaxies generally have elliptical morphologies and red optical colours, but have an excess of blue \nuvr\ colours, beyond that of old stellar populations \citep[][]{Kaviraj:2007aa, Schawinski:2007aa, Rawle:2008aa}. This blue NUV colour is thought to indicate the presence of a burst of recent star formation, most likely originating after a merger \citep[][]{Kaviraj:2007aa, Kaviraj:2011aa}, however there is little preference for location both within clusters, and in local density \citep[][]{Schawinski:2007aa, Yi:2011aa}. The origin of the young galaxy population is likely a merger induced burst of extra star formation, preceding a completely passive red sequence galaxy.

The second group of green valley galaxies are populations of spiral galaxies that are undergoing a slow decline in star formation \citep[][]{Schawinski:2014aa}. This population of slowly quenching spirals shows similarities to optically red, passive spiral galaxies \citep[e.g.][]{Wolf:2009aa, Bonne:2015aa, Fraser-McKelvie:2016aa}. These galaxies have spiral morphologies, but optical colours which suggest little widespread star formation \citep[][]{Masters:2010aa}. They also have lower star formation rates \citep[][]{Tojeiro:2013aa}, and appear in higher density regions than the blue, star forming counterparts \citep[][]{Skibba:2009aa, Bamford:2009aa}. They are found to still have residual NUV \citep[][]{Crossett:2014aa} and FUV fluxes \citep[][]{Moran:2006aa}, with star formation histories suggesting they may be the progenitors of large cluster S0 and elliptical galaxies \citep[][]{Mahajan:2009aa}. These red spiral galaxies likely form through interactions with group and cluster potentials, rather than with other galaxies, by truncating star formation without disturbing the typically fragile disc structure \citep[][]{Wolf:2009aa}. 

These results describe a transition from blue/star forming to red/passive that can occur in two broad ways; one dominated by fast morphological transformations and a quick cessation of star formation, and a slower process of mass growth and star formation decline, which may involve a group/cluster potential. The interaction between galaxy and environment therefore plays a key role in the evolution of its star formation. It is known that galaxies in denser regions have lower specific star formation rates \citep[e.g.][]{Lewis:2002aa, Gomez:2003aa}, redder colours \citep[][]{von-der-Linden:2010aa}, and generally more elliptical shapes \citep[][]{Dressler:1980aa} than those in less dense environments. These trends with density are more evident at small separations \citep[$<1$ Mpc,][]{Kauffmann:2004aa, Wilman:2010aa} typical of galaxy group scales. Galaxies in these more dense group environments also have their properties linked to those of their host central \citep[e.g.][]{Weinmann:2006aa, Prescott:2011aa}. The mechanisms in which groups can influence infalling members include: stripping the cold or warm gas supply in galaxies, \citep[][]{Gunn:1972aa, Quilis:2000aa, Balogh:2000aa}, tidal interactions and harassment between neighbours \citep[][]{Farouki:1980aa, Moore:1996aa}, and galaxy-galaxy mergers.

In this work we examine the processes occurring within these transition populations. To do so we look at optically red galaxies with blue \nuvr\ colours, such that the galaxy has a large old stellar population (from red optical colours), but an additional small population of young stars due to the presence of NUV \citep[][]{Yi:2005aa, Schawinski:2007aa}. These colour selections best isolate galaxies with low levels of recent star formation, with $\sim1\%$ of mass in young stars \citep[a residual amount of star formation, $<1M_{\astrosun}/\rm{year}$, ][]{Kaviraj:2007aa, Salim:2007aa}. 

If this \nuvr\ blue population of galaxies is indeed a result of the effects of group environment or neighbours, then looking at group properties would help disentangle the different pathways for these galaxies with residual star formation. The different group properties (central vs satellite, grouped vs ungrouped) can discriminate if falling into a larger halo does cause the excess of NUV flux.
We use the group catalogue of \citet[][]{Yang:2007aa} to look at the NUV in the process of a galaxy transitioning. In an idealised scenario, the quenching of a disc dominated system will require a long time-scale where the morphology does not change, whereas a burst of star formation is more likely merger driven. While these simplified scenarios do not account for extra gas infall onto ellipticals, or for any subtle effects of harassment/ram pressure, it can still be used as a basis for this investigation. 

In Section \ref{Section:Sample} we describe the Group catalogue by \citet[][]{Yang:2007aa}, and the \textsl{\textsc{Galex}} data employed for this study, as well as the use (and validity) of the selections made. In Section \ref{Section:Results} we present some primary results from this study, with \nuvr\ blue fractions used to compare the residual star formation within different populations of galaxies. In Section \ref{Section:Discussion} we discuss the roles of fast and slow quenching in different environmental samples as a way of explaining the results, before summarising in Section \ref{Section:Conclusion}. Throughout this work, we us AB magnitudes, and assume a flat cosmology with values of $H_0 = 71 {\rm km s^{-1} Mpc^{-1}}$ and $\Omega_M = 0.23.$

\section{Sample}
\label{Section:Sample}
We use the group catalogue of \citet[][]{Yang:2007aa} to provide group information for galaxies in the New York University Value-Added Galaxy Catalog \citep[NYU VAGC, ][]{Blanton:2005aa}. This catalogue improves on the halo-based group finder of \citet[][]{Yang:2005aa}, with an updated version for SDSS DR7 \citep[][]{Abazajian:2009aa}. We use only SDSS identified galaxies (to ensure the best matches with \textsl{\textsc{Galex}}) and model magnitudes (to avoid any systematic aperture biases from Petrosian magnitudes), and select galaxies between redshifts of $z=0.02$ and $z=0.05$ to match \citet[][]{Schawinski:2014aa}. For the group properties, we use a stellar mass limit of $\rm{log}_{10}(M_{star}) > 9.75$ which is brighter than the Petrosian magnitude limit of $r\sim 17.7$ and an absolute magnitude of $R\sim -20$ across the Sloan region. This limit is used for measuring group multiplicity and statistics, but is amended in the \textsl{\textsc{Galex}} matched sample to compensate for NUV completeness. The use of SDSS groups allows for a large overlap with NUV data, which we discuss below.

\subsection{\textsl{\textsc{GALEX}}}
\label{subsection:Galex}
We source our UV data from \textsl{\textsc{Galex}} General Release 5 \citep[][]{Martin:2005aa}, using the All-sky Imaging Survey (AIS) source catalogue from \citet[][]{Bianchi:2011aa}. This catalogue provides high quality \textsl{\textsc{Galex}} imaging from the inner 1\degree of the field of view of both NUV (1771--2831\AA) and FUV (1344--1786\AA), to a magnitude of $\sim20.8$ in the NUV and $\sim19.9$ in the FUV.
While catalogues in the Medium Imaging Survey (MIS) allow an extra magnitude in depth, we use the AIS for more spatially complete matching to SDSS sources. As we wish to maximise the amount of group members with a UV detection, using the MIS data results in many sources missing a UV counterpart due to the smaller footprint, giving potential false non-detections within groups. While limited MIS data is available for some sources, we find that using the mixed data does not have any significant effect on the results presented, so for simplicity and consistency, we use only AIS data of our sample. To overcome the shallow AIS data, we amend out mass limit for the UV matched data, which we explain below.

The final matched sample is adjusted to a limiting to a mass of  to ensure completeness of the NUV matches. For the UV properties of galaxies we use a stellar mass limited sample of $\rm{log}_{10}(M_{star}) > 10.35$. This results in a match rate of $90\%$ with the \textsl{\textsc{Galex}} catalogue of \citet[][]{Bianchi:2011aa} for galaxies within 3 arcseconds of the SDSS location, which are inside the survey area with good photometry in both NUV and FUV. Sources within the optical mass complete sample, but not in the UV matched sample (with a stellar mass of $10.35 > \rm{log}_{10}(M_{star}) > 9.75$ or without a matched UV counterpart) are included in the final matched sample for total group statistics, but not included in any NUV fraction. The higher mass limit ($\rm{log}_{10}(M_{star}) > 10.35$) in our UV matched sample ensures that the lowest mass galaxy within the UV matched sample will still have complete optical group statistics down to a factor of four smaller in mass. Additionally, the high mass limit provides a high match rate with \textsl{\textsc{Galex}},so the small percentage of galaxies without UV detections in AIS cannot change the \nuvr\ fractions in a meaningful way.

\subsection{Colour Selection and Corrections}
\label{subsection:Colours}
To ensure no redshift dependance on our colour results, we perform k-corrections from \citet[][]{Chilingarian:2010aa} to determine restframe \textsl{\textsc{Galex}} NUV magnitudes \citep[][]{Chilingarian:2012aa}. This method uses a polynomial fit based on a reference colour for each source and agrees with similar values from \textsc{KCorrect} code from \citet[][]{Blanton:2007aa}. We note that while the k-correction term is small ($\sim0.1$ magnitude), it is still calculated and included into the final sample. We correct for foreground dust extinction using $ugriz$ estimates from \citet[][]{Schlafly:2011aa}, and \textsl{\textsc{Galex}} NUV extinction estimates from \citet[][]{Yuan:2013aa}. To correct for internal dust extinction, Balmer line based corrections from \citet[][]{Oh:2011aa} are used. We use these $E(B-V)$ values with extinction curves of \citet[][]{Calzetti:2000aa} to model the attenuation for the SDSS photometry, and \citet[][]{Cardelli:1989aa} for the attenuation in the NUV, as per \citet[][]{Schawinski:2014aa}. The mean $E(B-V)$ value in the matched sample is $\sim0.07$ giving an attenuation correction of order $\sim0.5$ magnitudes in the NUV and $u$ band. All colours and magnitudes quoted are Balmer decrement dust corrected, and k-corrected to a redshift $z=0$, unless otherwise stated.

\begin{figure}
\begin{center}
\includegraphics[scale=0.55]{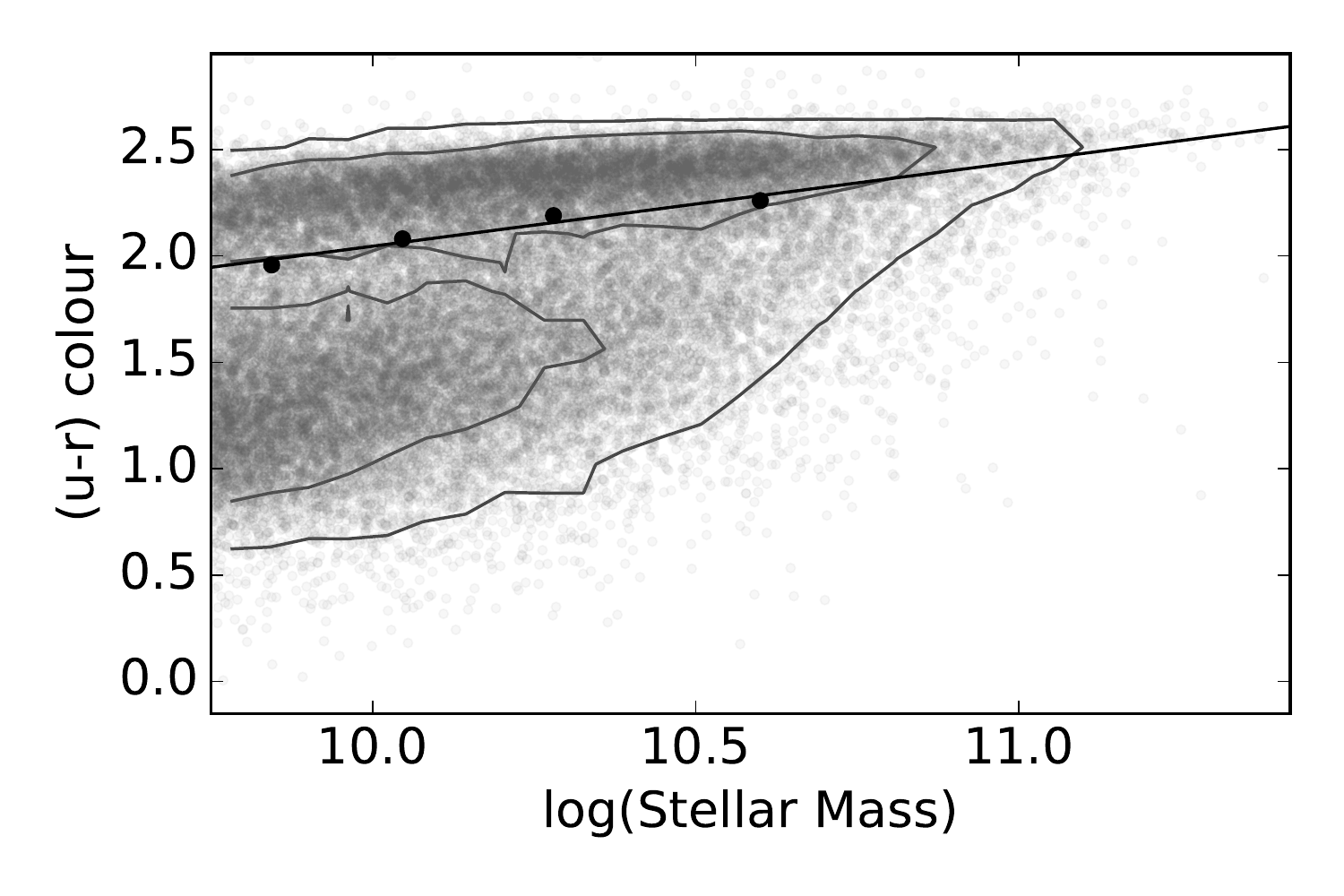}
\caption{Separation of red and blue sequences used for this study from a dust corrected \ur\ colour mass diagram. Black points represent the bisection of two Gaussians fitted to the colour distribution for different mass bins at the median stellar mass in each bin. The solid line represents the best-fitting line, which closely traces the density contour of the red population, confirming the accuracy of the fit.}
\label{fig:redsequences}
\end{center}
\end{figure}

In order to identify residual star formation in galaxies, we first select galaxies that reside on the red sequence. The optical red sequence has traditionally been used as a reliable way of selecting galaxies with older populations, and older mean stellar ages  \citep[e.g.][]{van-den-Bosch:2008aa, Peng:2010aa}. We use the \ur\ colour to select red sequence galaxies, to best straddle the 4000\AA\ break within our redshift range, and find the $u$ band uncertainties do not alter the shape of the distribution in a meaningful way.
To determine the population of red sequence galaxies, we take the distribution of k-corrected and dust corrected \ur\ colours in separate mass bins, and fit a double gaussian to the population, following a method similar to that of \citet[][]{Baldry:2004aa}. Briefly described, for stellar mass bins of equal number of objects, we take the bisection point between the two Gaussians as the delimiter for the populations, and use the resultant colour at the mean stellar mass of the bin to construct a linear separator of red and blue colour. Fig \ref{fig:redsequences} shows points at the mean stellar mass for four identified mass bins. The equation for the resultant linear colour separation is as follows:
\begin{equation}
 (u-r)= 0.239\times \rm{log}_{10}(M_{stellar})-0.154
\end{equation}
Where \ur\ is the k-corrected \ur\ colour at z$=0$, and $M_{stellar}$ is the stellar mass. We point out that only 4 bins are used here, however splitting the low mass bins to create additional points does not alter our distribution in any noticeable way. We see in Fig \ref{fig:redsequences} the separation line traces the 68\% contour of the red population, giving visual confirmation that this method has picked a sample of `red' colours, and older stellar populations.
Additionally, a maximum \ur\ colour of $(u-r)=3.0$ is also adopted to separate a low number of galaxies ($<10$) with potential photometric errors or high internal dust extinction, as well as unusually red objects that are not of interest in this study.

\begin{figure}
\begin{center}
\includegraphics[scale=0.55]{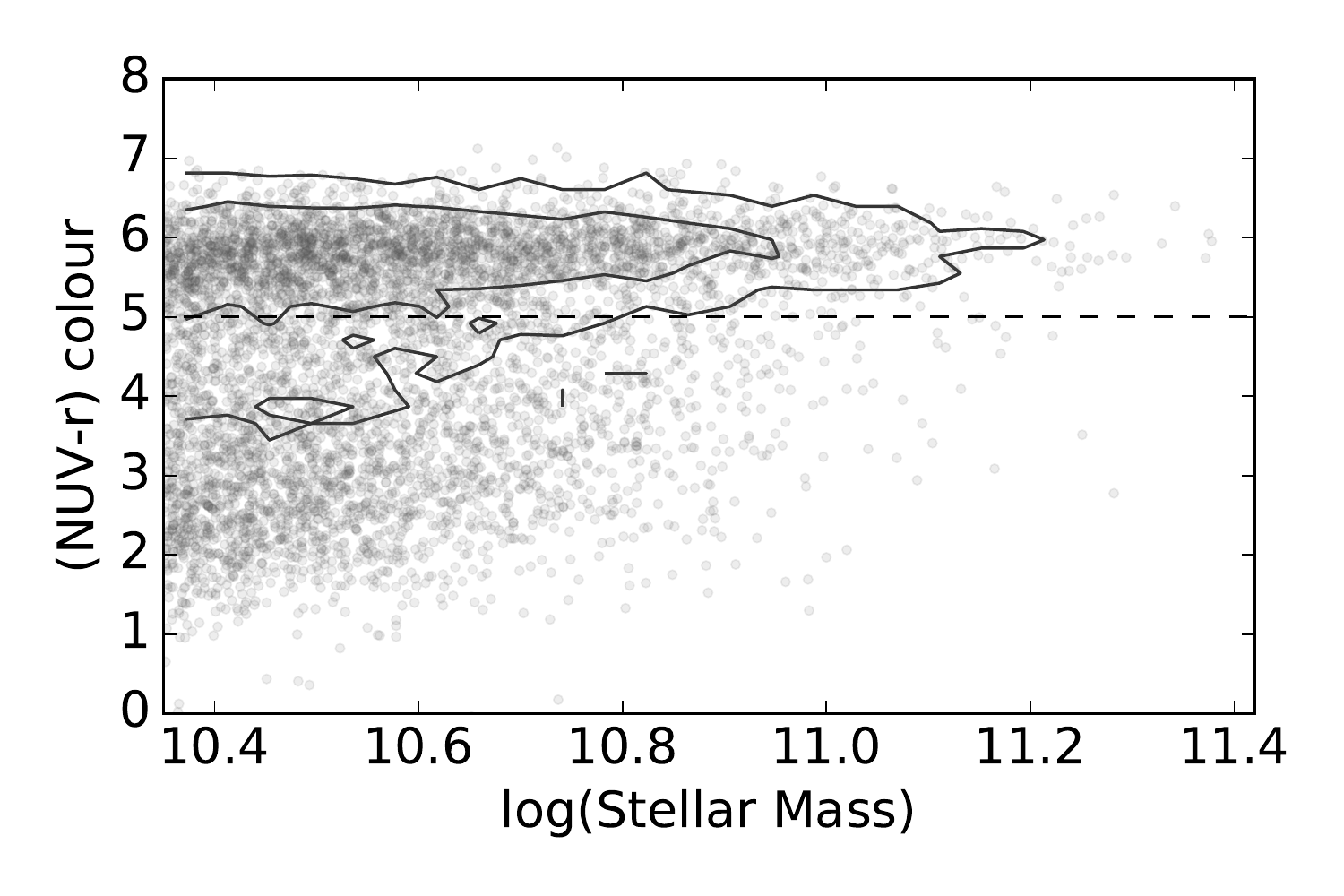}
\caption{The dust corrected \nuvr\ colour mass diagram for all galaxies in our matched sample. Contours denote red sequence galaxies defined in Fig \ref{fig:redsequences}. The extended tail of blue \nuvr\ galaxies suggests the presence of a young stellar population within these galaxies. The dashed horizontal line represents the threshold for residual star formation in this study.}
\label{fig:NUVdist}
\end{center}
\end{figure}

\section{Results}
\label{Section:Results}
In this section we look at residual star formation of galaxies within the red sequence population using the \nuvr\ colour mass diagram. In order to do this, we define an \nuvr\ colour that represents a residual amount of star formation. We choose the \nuvr\ colour to avoid significant contamination from UV upturn flux due to older stellar populations \citep[e.g.][]{Code:1979aa, Burstein:1988aa, OConnell:1999aa, Yi:2011aa}, which primarily affects the $({\rm FUV} - {\rm NUV})$ \citep[][]{Brown:2014aa}. 

Prior works use several different measures of \nuvr\ for residual star formation, with some studies defining maximum NUV contributions from old stellar populations at $\nuvr\ = 5.4$ \citep[e.g.][]{Kaviraj:2007aa, Schawinski:2007aa, Yi:2011aa}; while other works use a bluer limit for the contribution of young stars $\nuvr\ < 5.0$ \citep[][]{Jeong:2009aa}. \citet[][]{Salim:2007aa} and \citet[][]{Salim:2014aa} report star formation in the blue sequence at \nuvr\ $ < 4.0$, with a green valley $4.0 <$ \nuvr\ $ < 5.0$. We adopt a selection for residual star formation of \nuvr\ $ < 5.0$, to best represent galaxies with any residual star formation, while conservatively avoiding contributions from old stellar populations. The distribution of \nuvr\ as a function of mass in our sample is seen in the colour mass plot in Fig \ref{fig:NUVdist}.

This figure shows the \nuvr\ colour mass relation of all galaxies in the matched sample. Overlaid contours show \ur\ red sequence galaxies with a large red population, and a blue tail. We compute the fraction of red sequence galaxies with \nuvr\ $ < 5.0$, which would indicate the fraction of galaxies with residual are formation from NUV (hereafter \nuvr\ blue fraction), using our definitions from above. We find the \nuvr\ blue fraction to be $11.7 \%$ of our $\sim3000$ total red sequence sample (see Table \ref{tab:groupsdef} for details). This is below that of some previous results suggesting up to 30\% \citep[][]{Kaviraj:2007aa, Schawinski:2007aa}. The difference is due to selection effects of the chosen residual star formation threshold, which is more conservative compared to the former studies (see Section \ref{subsection:Colours}). Using an identical residual star formation threshold yields a fraction of $23 \pm 1\%$. This value is still less than previous studies, and is likely due to our sample being colour selected, while many previous studies using a light curve selection for their samples. Selecting galaxies based on red optical colours is likely to return a different fraction of red \nuvr\ galaxies, so is likely the cause of this discrepancy. 

\begin{figure}
\begin{center}
\includegraphics[width=0.5\textwidth]{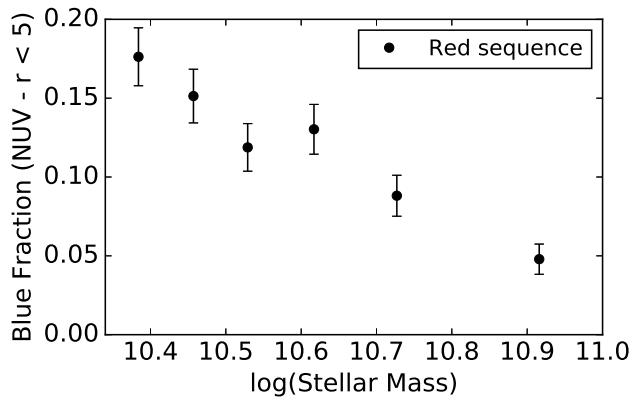}
\caption{NUV blue fraction for red sequence galaxies in different stellar mass bins. $1\sigma$ Poissonian uncertainties in each bin are also shown The residual star formation is largely ceased in most galaxies with higher mass in our sample, compared with the low mass galaxies.}
\label{fig:massdist}
\end{center}
\end{figure}

\begin{figure}
\begin{center}
\includegraphics[width=0.5\textwidth]{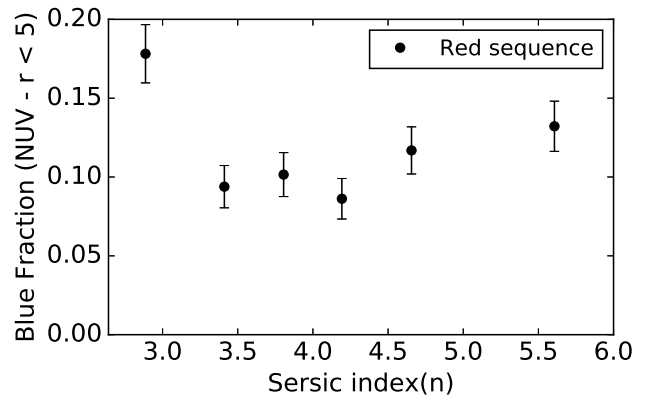}
\caption{NUV blue fraction for red sequence galaxies in different r band S\'ersic index bins. Low S\'ersic ($n<3$) galaxies show a somewhat higher fraction of \nuvr\ residual star formation than higher S\'ersic galaxies. This suggests more galaxies with discs and pseudobulges have residual star formation, compared to the bulge dominated galaxies.}
\label{fig:sercdist}
\end{center}
\end{figure}

Fig \ref{fig:massdist} shows the fraction of \nuvr\ blue galaxies binned by stellar mass. The blue NUV fraction is larger with lower mass galaxies, demonstrating as expected that mass does have an effect on the residual star formation of galaxies in our sample. The galaxies with higher mass have larger fractions that are completely passive, compared to low mass galaxies. We also find a modest rise the \nuvr\ blue fraction of lower r band S\'ersic index galaxies measured from the NYU VACG, seen in Fig \ref{fig:sercdist}. While a one component S\'ersic index is not a perfect measure of a galaxy morphology, it gives a reasonable measure of whether the light is dominated from a disc (such as in a spiral) or a bulge (seen in ellipticals). 

While this trend is seen for low S\'ersic index galaxies, the sample contains $< 1\%$ true disc-like (n$\sim1$) galaxies, with a median S\'ersic value of $\sim 2.85$ in the lowest bin. Despite this, we observe a modest increase in the \nuvr\ blue fraction of disc-like/pseudobulge galaxies  \citep[with S\'ersic index $n<2$, ][]{Fisher:2008aa, Ownsworth:2016aa}, compared to bulge-dominated galaxies ($n\sim4$). A comparison of our lowest S\'ersic index bin ($n<2.5$) to high S\'ersic galaxies ($n>3.8$) shows a change in fraction from $24 \pm 5\%$ to $10 \pm 1\%$, suggesting a preference for red sequence disc galaxies to have excess NUV flux. Using r band S\'ersic index values may result in more bulge dominated morphologies than using bluer bands \citep[][]{Kelvin:2012aa}, however variation in red galaxies is thought to be small \citep[][]{Vulcani:2014aa}. The variation is also usually higher in lower S\`ersic galaxies \citep[][]{Kelvin:2012aa}, such that any distinct disc features may not be present in the r-band image, that is being detected in \textsl{\textsc{Galex}}. This will likely make the results a lower limit on disc-like galaxies, which may be higher with more extensive S\`ersic profiling.

\subsection{Residual star formation and group environment}
In order to compare the group environment of galaxies, we construct definitions for galaxy environments. As well as requiring a definition of \textit{group} and \textit{non-group} galaxies, we also define samples within groups, based on their mass ranking within a group. We detail our definitions below, and use them throughout unless stated otherwise. Galaxies categorised as being grouped must have at least four other members in its halo to ensure accurate group information \citep[as per][]{Robotham:2011aa}, with groups sizes between two and four total members not considered in the following results. We further split the galaxies within groups as being \textit{central} or \textit{satellite}. Galaxies are considered central if they are the most massive galaxy in the group, and all other group members termed satellite. The numbers of galaxies are seen in Table \ref{tab:groupsdef}

We define galaxies as non-grouped when they are the central galaxy in its halo, with no other members above the mass limit within its host halo. This selection means that the lowest mass ungrouped galaxies ($\rm{log}_{10}(M_{star}) = 10.35$) will still have no companions within its halo down to a mass of $\rm{log}_{10}(M_{star}) = 9.75$. There may however, be smaller companions within the halo that have not been included in the mass complete sample. We assume the contribution of any undetected low mass galaxies affecting the non-group sample galaxies to be minimal however, given the lower mass limit to determine group multiplicity. These galaxies are still drawn from the group catalogue, so will not be completely isolated, but are known to not have any companions in their halo down to our mass limit. We therefore note the use of the word non-group, as distinct from isolated, which is not necessarily true of this population in the larger cosmic structure.

\begin{table}
\caption{Relative sample sizes and \nuvr\ blue fractions for each sample used in this study with Poissonian uncertainties. There is a higher percentage of non-grouped galaxies with \nuvr\ blue colours.}
\centering
\resizebox{0.5\textwidth}{!}{
\begin{tabular}{|c|c|cl}
\hline
Sample&Sample size& blue fraction (${\rm NUV} - r < 5.0$)\\
\hline
Total Red sequence & 3102 &$ 11.7 \pm 0.6 \%$\\
Central&205&$2.9 \pm  1.2 \%$\\
Satellite&734&$ 8.9 \pm  1.1 \%$\\
Total groups&839&$7.6 \pm  0.9 \%$\\
Non-group&1378&$16.2 \pm  1.1 \%$\\
\hline
\end{tabular}
}
\label{tab:groupsdef} 
\end{table}

\begin{figure}
\begin{center}
\includegraphics[width=0.5\textwidth]{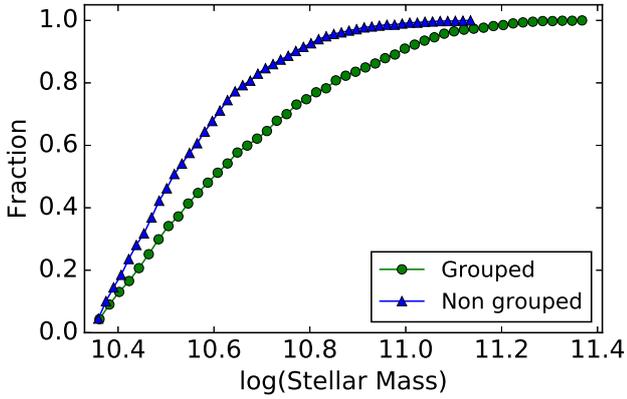}
\caption{Normalised cumulative distributions of the grouped (green, circles), and non-grouped (blue, triangles) galaxies stellar mass. A significant difference in the shape of the distribution is seen, with a K--S test indicating the samples are drawn from different distributions at  $> 4 \sigma$. This demonstrates that these samples are not drawn from the sample parent population, and stellar mass differences may be the cause of any differences between the samples.}
\label{fig:groupisol}
\end{center}
\end{figure}

\begin{figure}
\begin{center}
\includegraphics[width=0.5\textwidth]{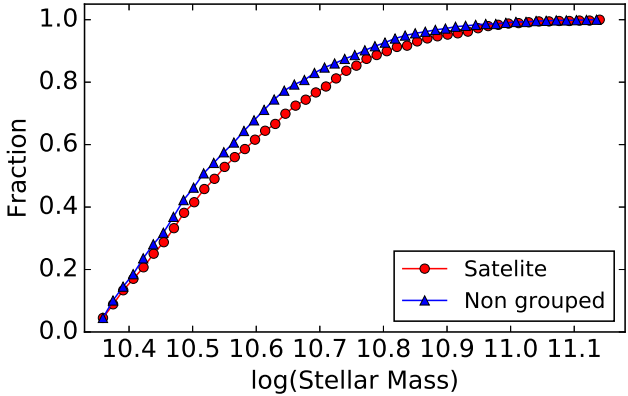}
\caption{Normalised cumulative distributions of the satellite (red, circles), and non-grouped (blue, triangles) galaxies stellar mass. Here the similarity is much higher, with a K--S test failed against the null hypothesis at $< 3 \sigma$. These samples are therefore less likely from two different distributions, such that differences between these samples are not due to mass.}
\label{fig:satisol}
\end{center}
\end{figure}

Using the previous definitions, we compute the fraction of galaxies in groups and non-grouped with $\nuvr\ < 5.0$ as $7.6 \%$, and $16.2 \%$ respectively. This suggests that within groups, the residual star formation is diminished compared to galaxies without massive halo companions. 
In Fig \ref{fig:groupisol} we show the mass distribution between the group and non-group sample. We note a systematic difference in both range and shape of the two mass distributions. As seen in Fig \ref{fig:groupisol}, the galaxy numbers per mass bin differ, leading to a systematic bias in the ungrouped sample, with an excess of lower mass galaxies. A Kolmogorov--Smirnov (K--S) test confirms that these cannot be drawn from a single population to a confidence of $> 4 \sigma$.
To better detect environmental differences without a mass bias, we reduce our grouped sample to only the satellite population. The fraction of galaxies with residual star formation is $8.9 \%$ in the satellite sample, still significantly less than the non-grouped sample. The central sample by comparison has a blue fraction of ($2.9 \%$) showing that the central sample is dominated by galaxies that are passive. 

We show in Fig \ref{fig:satisol} the cumulative mass distribution of satellite galaxies and non-grouped galaxies, showing a similar shape over the same mass range. The populations fail the K--S test against the null hypothesis at the $3 \sigma$ level such that they cannot be considered to be drawn from separate populations. A significant difference between the populations remains in the \nuvr\ colours, despite the similar mass profiles, with K--S test of \nuvr\ between the satellite and non-grouped galaxies indicating that the colours are not drawn from the same population at $ 4 \sigma$. We deduce that the residual star formation is quenched in galaxies in groups, compared to similar mass galaxies out of groups, due to the increased \nuvr\ fraction in our non-grouped sample, compared to the satellites.

\subsection{Environmental influence of residual star formation}
\begin{figure}
\begin{center}
\includegraphics[width=0.5\textwidth]{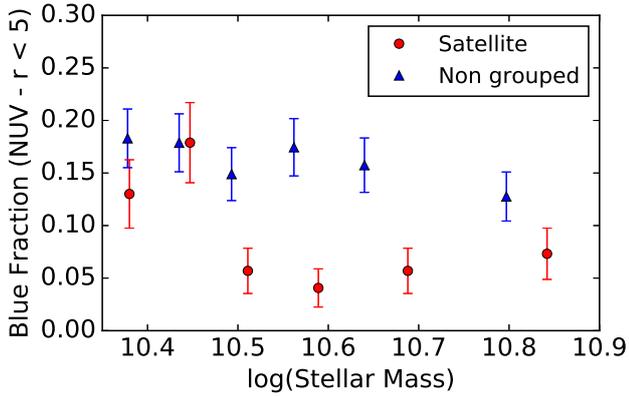}
\caption{The fraction of galaxies with residual star formation for different mass bins for both the satellite (red, circles), and non-grouped (blue, triangles) samples. There is a dependence on mass in the satellite galaxies, where the \nuvr\ blue fraction drops off above  $<10^{10.5}M_{\astrosun}$. This is not apparent in the non-group sample, and is almost flat over the mass range presented.}
\label{fig:masssatisol}
\end{center}
\end{figure}

\begin{figure}
\begin{center}
\includegraphics[width=0.5\textwidth]{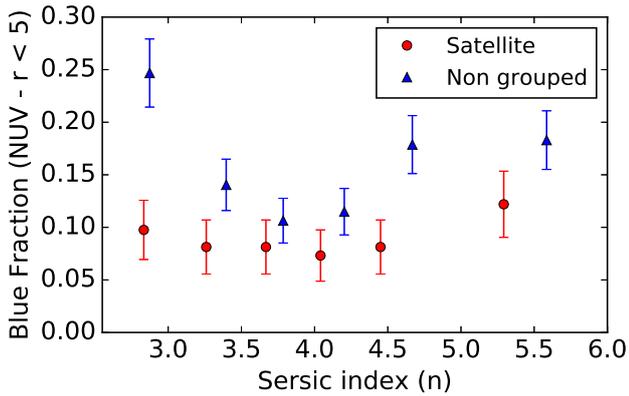}
\caption{The fraction of galaxies with residual star formation for different S\'ersic index n bins for both the satellite (red, circles), and non-grouped (blue, triangles) galaxies. While satellite galaxies do not show variation in the residual star formation, there is a higher proportion of non-grouped disc galaxies with residual star formation.}
\label{fig:Sersicsatisol}
\end{center}
\end{figure}

Here we explore in more detail the difference in residual star formation found between the group and the non-grouped environment. Fig \ref{fig:masssatisol} and \ref{fig:Sersicsatisol} show the \nuvr\ blue fraction for both the non-grouped and satellite galaxies, as a function of stellar mass and S\'ersic index respectively. Fig \ref{fig:masssatisol} shows a difference in the UV fraction for binned stellar mass, with a higher fraction of \nuvr\ blue galaxies with moderate stellar masses in the non-grouped sample. While both samples show a large \nuvr\ blue fraction, the satellite population decreases much more rapidly above $10^{10.5}M_{\astrosun}$. In contrast, the non-grouped sample remains almost flat throughout the entire mass range, dropping slightly at higher masses where there is no distinguishable difference between samples. This suggests that the group environment has an effect on quenching galaxies at higher masses, which is not seen in non-grouped galaxies.

Fig \ref{fig:Sersicsatisol} shows the NUV fraction for the same two samples, binned by S\'ersic index $n$. We see the \nuvr\ blue fraction does not have the sample shape between the two samples. We see that the \nuvr\ blue fraction is constant in the satellite sample across the entire S\'ersic range. The non-group sample however is not constant, but has higher blue fraction in the lowest S\'ersic bin. This higher NUV fraction occurs at S\'ersic indices that are generally associated with light dominated by pseudobulges \citep[$n\sim2$][]{Andredakis:1994aa, Kormendy:2004aa} and spiral type ($n<2$) morphologies. We suggest the increase in \nuvr\ blue fraction in the non-group sample may be due to additional low S\'ersic galaxies with blue \nuvr\ colours seen in the lowest S\'ersic bin, compared to the satellite sample. Taken with the previous result, we see that galaxies outside of groups cease star formation in a longer time, allowing residual star formation at higher masses. This process would also be less disruptive to galaxy structure, with more disc dominated morphologies seen outside of groups.

\subsection{Comparisons to stellar population models}
To further investigate the possibility of the formation of the \nuvr\ blue population within and outside groups, we look at the evolution of star formation in a galaxy in the \ur\ \nuvr\ plane. We do this with the use of stellar population models with various star formation histories, and compare the evolutionary tracks with the data from our samples. \ur\ against \nuvr\ highlights differences between the mean stellar age represented by optical colours, and star formation traced by $(\rm{NUV}-\rm{r})$, akin to that of \citet[][]{Schawinski:2014aa}. If star formation is shut off slowly, then the \nuvr\, which is sensitive to low levels of star formation, may still appear blue due to the presence of young stars. The optical colours however, which are more affected by the overall stellar population, would become red. Conversely, if star formation were to be completely shut off instantaneously, then the \nuvr\ colour should turn red with the optical colour.

\begin{figure}[ht]
\begin{center}
\includegraphics[width=0.5\textwidth]{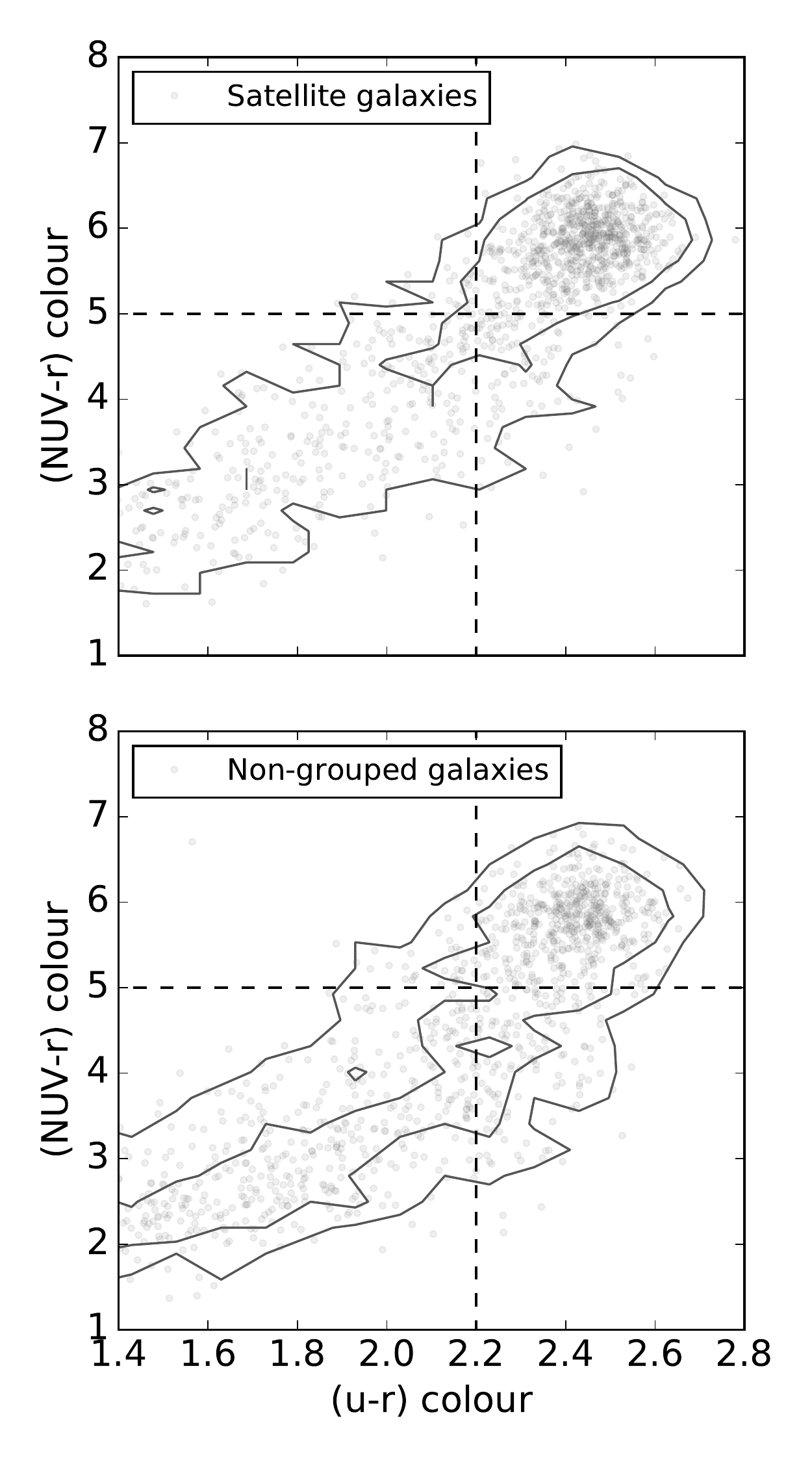}
\caption{ \ur\ -- \nuvr\ Colour-colour diagram for the satellite (top) and non-grouped (bottom) samples. The contours denote the 68, and 95\% levels, with the non-grouped sample having displayed only a random selection of points to match the number in the satellite sample. Dashed lines denote the UV residual star formation threshold $( \nuvr\ = 5.0)$, and  \ur\ optical red sequence threshold, for the minimum mass galaxy in the sample $(\ur\ = 2.2)$. The quadrants therefore represent different general photometric properties, with blue galaxies in the lower left hand quadrant, traditional red sequence galaxies in the upper right, and the \nuvr\ residual star formers in the lower right quadrant.}
\label{fig:doubleplot}
\end{center}
\end{figure}

Fig \ref{fig:doubleplot} shows the \ur\ and \nuvr\ colours of  all galaxies in both satellite and non-grouped samples, with contours at the 68, and 95\% levels. Overlaid are dashed lines to denote the UV residual star formation threshold ($\rm{NUV}-\rm{r}= 5$), and the \ur\ optical red sequence for the minimum mass galaxy in the sample ($\rm{u}-\rm{r} = 2.2$). Galaxies that are not on the red sequence are also included in the figure for illustrative purposes. In particular, Fig \ref{fig:doubleplot} (bottom) shows the non-grouped galaxy sample has an increased population in the bottom right of the plot, and higher diversity of blue \nuvr\ colours, with a tenth percentile of \nuvr\ $=4.6$ and a variance of $\sigma_{NUV}^2 \sim 0.5$ for the red sequence. In contrast, the satellite galaxies in Fig \ref{fig:doubleplot} (top), show comparatively smaller diversity, with much more condensed contours in the main population in the upper right. The sample displays a tenth percentile \nuvr\ colour of the red sequence galaxies are \nuvr\ $=5.1$ with a variance of $\sigma_{NUV}^2 \sim 0.3$. This suggests that not only is there a higher number of \nuvr\ blue galaxies outside of groups, but they also have bluer tenth percentile colours and small increase in variance.

\begin{figure*}[ht]
\begin{center}
\includegraphics[width=\textwidth]{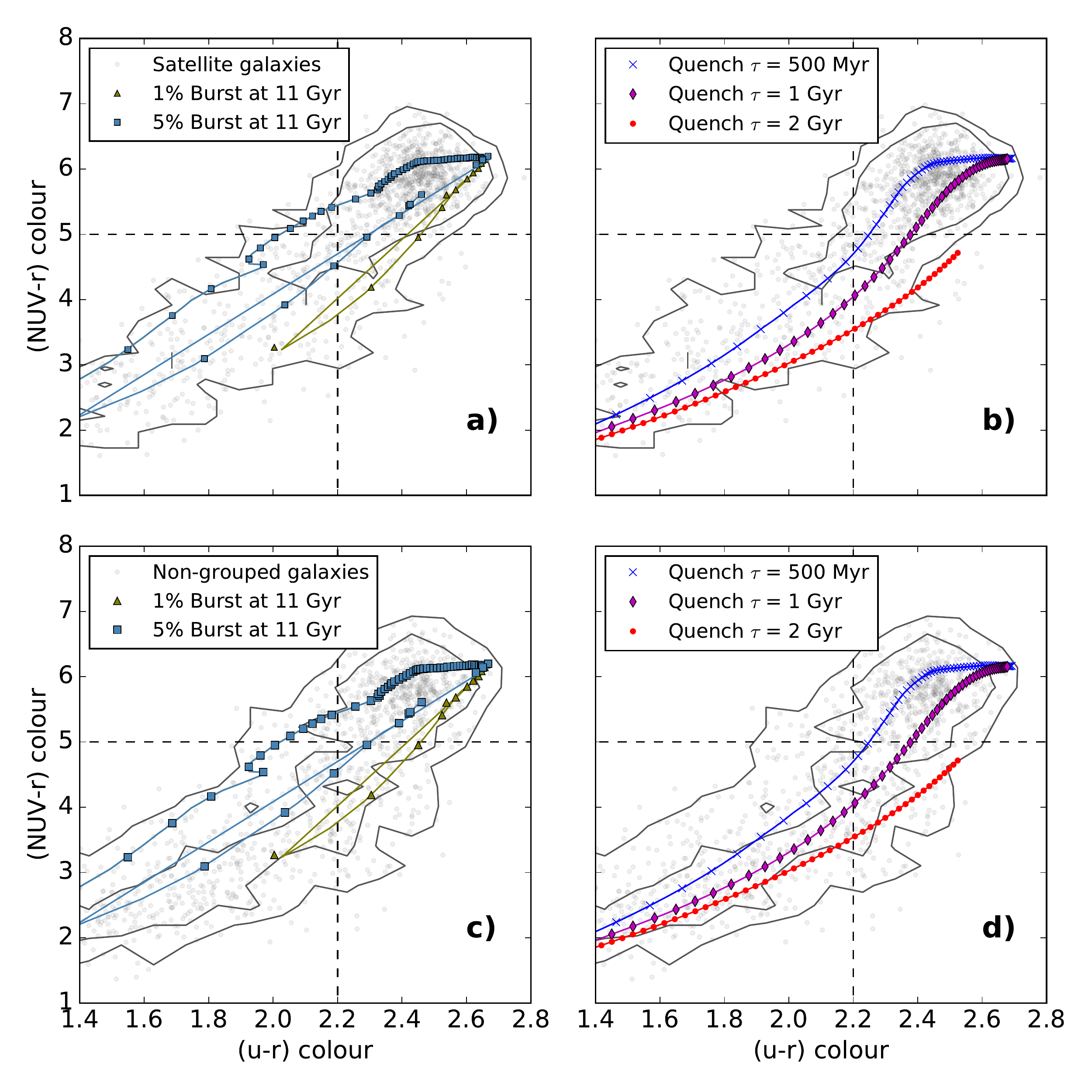}
\caption{Same as Fig \ref{fig:doubleplot} with the inclusion of various stellar population models from \citet[][]{Bruzual:2003aa}. Coloured lines represent stellar population tracks with different star formation histories, with points depicting spline interpolated time steps of $\sim0.2$ Gyr. The stellar population synthesis models in panels (a) and (c) contain a single old stellar population, with an extra burst of formation at a 1\% (yellow, triangles) and 5\% (blue, squares) relative strength at  $t_{present}-2$ Gyr. Panels b and d show exponentially decaying star formation rate tracks overlaid, with e-folding times of $\tau = 500$ Myr (navy, crosses), $\tau = 1$ Gyr (magenta, diamonds), and $\tau = 2$ Gyr (red, circles). Burst models only briefly move through the \ur\ red, and \nuvr\ blue quadrant, compared to the decay models, indicating a preference for decays over bursts. Additionally, longer $\tau$ models better match the \nuvr\ colours of the non-grouped sample, compared to the satellite sample, where long decays avoid the majority of sources.}
\label{fig:multiplot}
\end{center}
\end{figure*}

In Fig \ref{fig:multiplot} we overlay various stellar population tracks from \citet[][]{Bruzual:2003aa} onto these contours, to better establish the possible differences in quenching from the colour-colour diagram. This figure shows models overlaid on the satellite and non-grouped samples, with burst models in the left hand panels, and decay models in the right hand panels, with interpolated time steps of $\sim0.2$ Gyr. A \citet[][]{Chabrier:2003aa} initial mass function (IMF) is used for all models, and a Solar metallicity. Models using a \citet[][]{Salpeter:1955aa} IMF do not affect the result in a significant way. We find changes to the metallicity cause the tracks to remain blue in \ur\ \citep[as in ][]{Rawle:2008aa}, and thus are not considered for this analysis. The colours used are based off model rest frame magnitudes.

The burst models in the left hand panels contain a single burst at $t=0$, and then a smaller burst of 1\% (yellow, triangles) and 5\% (blue, squares) at  $t_{present}-2$ Gyr. We see that these models cover the majority of galaxies in the satellite sample (panel a), reaching the 95\% contour boundary for the blue \nuvr\ galaxies. However, the galaxies do not have the excess \nuvr\ colour for very long ($<0.5$ Gyr),  making this kind of transition relatively uncommon. In addition to this, panel (c) of Fig \ref{fig:multiplot} shows the burst models do not successfully reproduce the diversity of colours seen in this population. We therefore assume that while the burst may explain the colours of some \nuvr\ galaxies, it cannot be the dominant cause of the \nuvr\ blue population.

As an alternative model, we use exponentially declining star formation rates to best simulate cessation of star formation. Fig \ref{fig:multiplot} (b, d) shows the satellite and non-grouped contours respectively, this time with the addition of three different quenching events. The characteristic e-folding time-scale of each of these is 500 Myr (navy, crosses), 1 Gyr (magenta, diamonds), and 2 Gyr (red, circles). 

Here we see in panel (b) the fast decaying model explains almost all of the \nuvr\ blue galaxies with longer decays actually missing the majority of points presented. By contrast, the extended scatter in panel (d) demonstrates that longer ($>1$ Gyr) decays are also present in this non-grouped sample. In Fig \ref{fig:multiplot} (b) the $\tau = 2$ Gyr track traces close to the 95\% percentile contour and contains very few points that can be explained by this model, with only 6\% of points bluer than the long decay model. By contrast, the 2 Gyr model traces much closer to the 68\% contour for the non-group sample, with 14\% of points bluer than this curve. This shows that the diverse \nuvr\ colours in the non-group sample are more accurately represented by a combination of long and short $\tau$ models, compared to the satellite sample. This leads to the conclusion that a larger percentage of the non-grouped, \nuvr\ blue galaxies are undergoing a long time scale decay of star formation, which is not prevalent in the satellite sample.

\section{Discussion}
\label{Section:Discussion}
In the previous sections, we have investigated the residual star formation of red sequence galaxies within groups from SDSS, and whether differences in group and non-group populations infer different mechanisms that cause the presence of star formation at the $\sim1\%$ by mass level. We have used \nuvr\ blue fractions to quantify the amount of residual star formation in our sample, finding the fraction to be $\sim 12\%$ of our total red sequence sample. 

We find there is a significant difference in the residual star formation between group and non-group galaxies, with the latter containing a fraction of galaxies with \nuvr\ residual star formation of $\sim 16 \%$, compared to satellite galaxies with only $\sim 9\%$. Previous studies have found little density change for \nuvr\ fraction \citep[e.g. ][]{Schawinski:2007aa, Yi:2011aa}, but do not specifically probe group halo environment, preferring a local density measure instead. The results do broadly match the idea of environment quenching of star formation \citep[e.g. ][]{Peng:2010aa}, with colour fractions inside and outside groups similar to that of \citet[][]{van-den-Bosch:2008aa}.

A higher blue fraction of comparatively low ($n<2.5$) S\'ersic index red sequence galaxies is also seen outside of groups, compared to satellites. These galaxies may form part of the class of red spirals seen elsewhere \citep[e.g. ][]{Masters:2010aa, Bonne:2015aa} with the red optical colours caused by the large underlying old stellar population \citep[][]{Cortese:2012aa}. Red spirals are known to have higher star formation rates than similar mass red ellipticals \citep[][]{Tojeiro:2013aa} so are more likely to contain residual star formation, and thus are a potential candidate for our low S\`ersic population. 

The scenario presented matches that of \citet[][]{Schawinski:2014aa}, whereby disc-like galaxies move onto the red sequence in a different pathway to elliptical galaxies. The smaller residual star formation fraction for both disc galaxies and high mass galaxies inside groups, compared to those outside of groups indicates quenching is likely to have occurred early on in the mass evolution due to environment. The non-grouped galaxies will move onto the red sequence later in their mass evolution, and retain lower S\'ersic indices at higher mass. 

We use \ur\ and \nuvr\ colour-colour diagrams with \citet[][]{Bruzual:2003aa} stellar population models, to test different formation scenarios of the blue \nuvr\ colour. These tracks are analogous to those of \citet[][]{Schawinski:2014aa} and \citet[][]{Smethurst:2015aa}, who found longer decay timescales to be more common in morphologically selected spiral galaxies. Our study finds a similar result, that galaxies undergoing these slower quenches are preferentially non-grouped, and have low S\'ersic indices. Our satellite sample has an evolution similar to that of \citet[][]{Wetzel:2013aa}, where galaxies once falling into groups have a short ($<0.8$ Gyr) quench time, however differs from that of \citet[][]{Skibba:2009ab}, whose models show longer quenches are preferred in groups, taking  $>2$ Gyr. The model design however, dictates that these galaxies move onto the red sequence at high mass. This differs from our red sequence satellites, which show a lower fraction of residual star formers at high mass.

These results indicate that the excess of residual star formation outside group environments must be due to a long, slow quench in star formation. In contrast, the same stellar population tracks do not represent the majority of the satellite galaxy population. These galaxies have instead fully quenched, which has stopped any residual star formation from occurring. This shows that star formation on residual levels can be due to environmental effects, such that residual star formation in galaxies within groups will be quenched on short timescales, compared to those outside. 

\citet{Rasmussen:2012aa} hypothesize that star formation quenching in galaxies from group accretion would be larger than the typical crossing time of a group ($1.1-1.6$ Gyr), suggesting a longer timescale quenching than our results. This finding agrees with simulations of ram pressure in group environments from \citet{Steinhauser:2016aa} showing the migration of galaxies on the \ur\ \nuvu\ colour-colour diagram from blue to red would take $> 1$ Gyr due to ram pressure, consistent with other ram pressure models \citep[e.g. ][]{Tonnesen:2007aa}. \citet{Rasmussen:2012aa} suggests that tidal interactions between galaxies could speed up this quenching process, and would present a short lived enhancement of star formation. Our satellite sample is therefore likely quenching due to a combination of these processes, with ram pressure alone insufficient. However, our red sequence sample eliminates galaxies many star formation enhancement signatures to test this theory further.

Our results demonstrate that the blue \nuvr\ colours of our group galaxies are best explained solely by the rapid quenching of galaxies. Mechanisms such as the `strangulation' process, where the group/cluster medium starves a galaxy of gas and star formation on scales of several Gyr \citep[][]{Larson:1980aa, Balogh:2000aa, Peng:2015aa}, are therefore not the dominant mode of residual star formation quenching of our satellite sample. Instead, processes such as ram pressure stripping \citep[][]{Gunn:1972aa} that act on timescales of  $<500$ Myr \citep[][]{Bahe:2015aa}, appear more likely to cause the decline in star formation.

The spread of S\'ersic indices suggest that morphology may also be altered. S\'ersic profiles of post-merger galaxies are generally higher \citep[n$\sim4$, ][]{Aceves:2006aa}, with processes such as tidal interactions \citep[][]{Barnes:1991aa} and harassment \citep[][]{Moore:1996aa} also potential candidates for the truncation of the grouped galaxies. A merger induced burst of a red galaxy is also possible instead of a quench, but models in Fig \ref{fig:multiplot} (a) suggest this is transition is short-lived, so a less likely candidate. The lack of slow quenching provides more evidence of the importance of a group environment on the rapid transformation of star forming galaxies once truncation starts \citep[][]{Wetzel:2012aa, Wetzel:2013aa}.

By contrast, the non-grouped sample shows colours consistent with a mix of rapid and slow quenching events. Many of the non-grouped galaxies are thus likely undergoing a secular ($>1$ Gyr timescale) form of evolution. \citet[][]{Pan:2014aa} found that late-type green valley galaxies with \nuvr\ blue disks are likely to have formed from secular processes, matching this result. Many of the excess galaxies are therefore likely to be secularly evolving, disk and pseudobulge galaxies, with secular processes common in pseudobulge galaxies \citep[][]{Kormendy:2004aa}, and pseudobulge galaxies known to have red optical colours \citep[][]{Fernandez-Lorenzo:2014aa}.

While many of the galaxies in the non-group sample are best suited to a long decay model, other sources have colours that can be traced by a fast decay. These could be the result of the larger scale environment, with processes acting on the galaxy outside the local halo environment \citep[e.g. ][]{Bahe:2013aa}. These processes are not able to be looked at without knowledge of the wider structure, so are left unknown in this study. Another possible contaminant in the non-grouped sample are leftover remnants of fossil groups \citep[e.g. ][]{Ponman:1994aa, Jones:2003aa}. These galaxies would have undergone a rapid quench, or a merger induced burst of star formation, leaving the result appearing to be alone within a single halo, but still a result of direct environmental processes. Further analysis of the haloes could help isolate this possible scenario, but is left as a future endeavour.

\section{Summary}
\label{Section:Conclusion}
In this paper we have taken the SDSS DR 7 group catalogue from \citet[][]{Yang:2007aa}, in conjunction with all-sky data from \textsl{\textsc{Galex}} GR5 \citep[][]{Bianchi:2011aa} to investigate the residual star formation of galaxies in groups. Our main findings are:

\begin{itemize}
\item In red sequence galaxies, residual star formation from \nuvr\ colour is seen in $\sim12\%$ of our mass limited sample. Galaxies with this residual star formation are predominantly low mass ($<10^{10.5}M_{\astrosun}$) galaxies, with a small excess of comparatively lower S\'ersic indices ($n<3$).

\item Residual star formation is preferentially seen in non-grouped galaxies, compared to either satellites or central galaxies within groups, with the difference in \nuvr\ between non-grouped and satellite populations due to an environmental effect.

\item The residual star formation fraction falls by a factor of three in satellite galaxies above $10^{10.5}M_{\astrosun}$, but remains largely constant ($>10\%$) in non-grouped galaxies across the entire mass range. Additionally, there is a higher fraction of low S\'ersic ($n<3$) galaxies in the non-grouped sample, compared to satellites. 

\item The diversity of \nuvr\ colours in the non-grouped sample suggests that many galaxies are best matched to long ($\sim 2$ Gyr) decay times using models from \citet[][]{Bruzual:2003aa}. The satellite population can be represented by solely short decay models ($\sim500$ Myr).
\end{itemize}

We suggest that taken together, many galaxies outside of groups are preferentially undergoing a slow quench, retaining discy morphologies implying a secular form of evolution, where small amounts of star formation continue to occur for up to $\sim2$ Gyr. By contrast, satellite galaxies have shorter quenching timescales, that fully quench residual star formation within $<1$ Gyr, which would likely suggest a combination of tidal interactions and ram pressure to quench star formation in such a short time. This demonstrates that environment is important for the quenching of even the last remnants of star formation, whereas galaxies isolated from groups are able to form low amounts of new stars for longer times. Information regarding galaxy environment is therefore an important factor in knowing the star formation history of a galaxy.

\section*{Acknowledgements}
The authors thank the anonymous referees for their feedback which greatly improved the overall quality of this work. The authors also thank Sugata Kaviraj for providing the group catalogue used in this study. JPC acknowledges financial support through the Australian Postgraduate Award and thanks the University of Hull for hospitality while much of the work was completed.  
MJIB acknowledges financial support from The Australian Research Council (FT100100280) and the Monash Research Accelerator Program (MRA).
JPS gratefully acknowledges support from a Hintze research fellowship.

Funding for the SDSS and SDSS-II has been provided by the Alfred P. Sloan Foundation, the Participating Institutions, the National Science Foundation, the U.S. Department of Energy, the National Aeronautics and Space Administration, the Japanese Monbukagakusho, the Max Planck Society, and the Higher Education Funding Council for England. The SDSS Web Site is http://www.sdss.org/.

The SDSS is managed by the Astrophysical Research Consortium for the Participating Institutions. The Participating Institutions are the American Museum of Natural History, Astrophysical Institute Potsdam, University of Basel, University of Cambridge, Case Western Reserve University, University of Chicago, Drexel University, Fermilab, the Institute for Advanced Study, the Japan Participation Group, Johns Hopkins University, the Joint Institute for Nuclear Astrophysics, the Kavli Institute for Particle Astrophysics and Cosmology, the Korean Scientist Group, the Chinese Academy of Sciences (LAMOST), Los Alamos National Laboratory, the Max-Planck-Institute for Astronomy (MPIA), the Max-Planck-Institute for Astrophysics (MPA), New Mexico State University, Ohio State University, University of Pittsburgh, University of Portsmouth, Princeton University, the United States Naval Observatory, and the University of Washington.

This work has made use of public \textsl{\textsc{Galex}} data. \textsl{\textsc{Galex}} is a NASA Small Explorer, launched in 2003 April. We gratefully acknowledge NASA's support for construction, operation and science analysis for the \textsl{\textsc{Galex}} mission, developed in cooperation with the Centre National d'Etudes Spatiales of France and the Korean Ministry of Science and Technology.

\bibliographystyle{mnras}
\bibliography{V5refs}
\end{document}